\documentclass[12pt,a4paper,final]{iopart}

\usepackage{iopams}
\usepackage{graphicx}
\usepackage{comment}
\usepackage[breaklinks=true,colorlinks=true,linkcolor=blue,urlcolor=blue,citecolor=blue]{hyperref}
\usepackage{siunitx}
\expandafter\let\csname equation*\endcsname\relax
\expandafter\let\csname endequation*\endcsname\relax
\usepackage{amsmath}

\newcommand{\ts}{\textsuperscript}

\begin{document}

\title{The IYPT and the Ring Oiler Problem}

\author[cor1]{Martin Plesch$^{1,2}$}
\address{$^1$Institute of Physics, Slovak Academy of Sciences, Bratislava, Slovakia}
\address{$^2$Institute of Computer Science, Masaryk University, Brno, Czech Republic}
\ead{martin.plesch@savba.sk}

\author{Samuel Plesn\'ik}
\address{Department of Physics and Astronomy, University College London, United Kingdom}

\author{Nat\'alia Ru\v zi\v ckov\'a}
\address{Institute of Science and Technology Austria, Klosterneuburg, Austria}

\begin{abstract}
The International Young Physicists' Tournament (IYPT) continued in 2018 in Beijing, China and 2019 in Warsaw, Poland with its 31\ts{st} and 32\ts{nd} editions. IYPT is a modern scientific competition for teams of high school students, also known as the Physics World Cup. It involves long time theoretical and experimental work focused on solving 17 publicly announced open ended problems in a teams of five. On top of that, teams have to present their solutions in front of other teams and scientific jury and get opposed and reviewed by their peers. Here we present a brief information about the competition with a specific focus on one of the IYPT 2018 tasks - the Ring Oiler. This seemingly simple mechanical problem appeared to be of such a complexity that even the dozens of participating teams and jurying scientists were not able to solve all of its subtleties.

\end{abstract}

\vspace{2pc}

\section{Introduction}

Modern educational methods give more and more emphasis on an independent inquiry work of the students. Rather than passively receiving knowledge from the lecturer and later delivering it back during tests and exams, students are expected to develop their knowledge and understanding by themselves under an expert supervision. This is often done in teams, where different students can focus on different aspects of the task and join their forces to achieve a common target. 

The International Young Physicists' Tournament  \cite{iypt.org, iypt2016, iypt2017} unites all these modern education aspects. Each summer, $17$ open-ended physics problems are published by the International Organising Committee and all students worldwide are welcome to start their work on finding suitable solutions. They work in teams of five (in many countries on different levels of the competition this number is smaller, up to involving individuals as well) and are expected to seek for available information on the tasks, consult experts, develop theories and perform experiments. 

During the competition itself, students are left on their own with their peers. There is no lecturer to provide and correct tests and tasks, there is no examiner to ask tough questions. Teams meet in groups of three (four in some cases) and are the central point of the competition for almost the whole time. One of the teams, the opposing team, challenges the reporting team for one of the problems. The challenge might be rejected (as no-one knows everything), but too many rejections cause a loss of points. If the problem is accepted, the reporting team presents its solution and defends it against the opposing team that naturally tries to find any flaws and drawbacks. Close to the end of the stage, the third team comes into play, providing a brief review of the stage, highlighting positives and negatives of both the reporter and the opponent. 

There is a jury, of course. The chair helps to organise the run of the stage, announces the start and the end of each team's performance or preparation. Nevertheless, in most cases, after introducing the jury members and the teams at the very beginning he or she can let the initiative to the teams until the very end of the stage. A few minutes stand for questions from the jury members, which are, however, not meant to test the knowledge or ability of the students. Rather than that, they should aim to explain anything that has been unclear to the jurors so far. After that, the jurors independently show their grades ranging from 1 to 10 for all three teams in a figure-skating style and, very importantly, have to justify their grade if it was on the edge of the distribution. 

The more open and interesting the tournament is for students, the more demanding it is for jurors. Growing complexity of the tasks and its solutions together with the large spread of the fields the tasks cover makes it harder and harder for the jurors each year. Within less than an hour, the juror has to listen to the solution, a lot of criticism and refute, finally a more or less objective review and decide on the grade within a few minutes. 
This is why jurors are led by a complex scoring system which aims to give a guide to finding the final score by allotting partial scores to different parts of the performance of the teams.  

The open and democratic attitude of IYPT towards students is underlined by the fact that not only jurors award grades to the teams, but also teams are invited to provide feedback on the qualities of the chair and the jurors. One might expect that this feedback handed in by the teams would be strongly correlated with the graded the team has received, but interestingly, this is not the case. While students can more or less happily accept strict grading, if duly justified, they tend to (rightfully) criticise poor questions from the jurors or groundless high scores given to their peers.  

\section {The 31\ts{st} IYPT 2018 in Beijing, China}
The 31\ts{st} IYPT was hosted by one of the best universities in China, the Renmin University in Beijing. As it happens, the preparation of the tournament was not saved from complications; tensions between China and western countries combined with the difficulty in cooperation between local authorities and a foreign NGO nearly resulted in the tournament being cancelled. 
But finally, thanks to the unified effort of the IYPT management and the local organisers, we have had a tournament prepared like very few before: excellent competition rooms, services, food, accommodation and excursions. The level set by Chinese organisers will be very hard to beat in the future. 

In spite of the uncertainties during the preparations, the number of participating countries continued to grow. 32 teams made it a bit hard for the organisers with two groups of four teams, but the joy of having more and more excellent students on the board was certainly stronger. All of them worked on the same set of 17 problems published after the end of IYPT 2017 \cite{problems2018}.

Among the most interesting ones was the \textit{Ring Oiler} problem, which will be discussed in detail in the Section \ref{Oiler}. Many of the teams got overwhelmed by the \textit{Drinking straw} task as well. Probably everyone has already experienced that in certain cases a straw will rise from a fizzy drink and potentially topple. However, stating the exact conditions for this phenomenon to occur is far from trivial. With too little drink in the glass, the straw will not rise high enough to topple, but for a fully filled glass it will stick on the top in a horizontal position due to surface tension and will not topple either. Keeping the level of carbonisation of the drink constant during the experiment turned to be also a tricky issue.   

\textit{Candle in the water}, not in the wind, was the next interesting problem. Different effects can cause a barely floating candle with a weight attached float. Some of the candles will form a cavity of air inside that will help floating due to Archimedes force, others will achieve the same effect by pouring the melted wax on the surface of water and get the support by surface tension. And others will extinguish very soon. 

The problem of \textit{Tesla valve} was a rather different type of task. Students were asked to experimentally recover the old patent of Nikola Tesla: a one-way valve for fluids. This passive element was originally intended to be useful in steam engines and power plants, but it turned out that the effect was smaller than expected. With current technology of finite element simulation and well as 3D printing it is much simpler to do both theoretical and experimental investigation. We have seen a lot of interesting results proving the fact that the asymmetry in flow seems to be smaller than Tesla expected.

Levitation is certainly a fascinating phenomenon and the problem \textit{Acoustic levitation} asked students to design and test a device that would be able to levitate small object using acoustic standing waves. This was particularly exciting for young engineers who sometimes used more than a dozen of sound sources for a precise manipulation of small pieces of polystyrene. And yet, sometimes the core question -- why exactly the object levitates -- has not been completely cleared out in the discussion between the reporter and the opponent. 

And last, but not least, the \textit{Water bottle} task asked the students to describe how to throw a partially filled bottle in such a way that it performs a somersault and lands in an upright position. While some teams engaged skilled colleagues for performing the experiments, other designed sophisticated launching robots. The trick turned out to be to fill the bottle close to the point where the centre of mass of the bottle and water is the lowest (depends on the weight/volume ratio of the bottle) and throw it in such a way that gravitational force will overwhelm the centrifugal force on the water in the bottle and make the water flow through the bottle shortly before landing.  

\section {The 32\ts{nd} IYPT 2019 in Warsaw, Poland}
A tournament in Europe usually promises a large number of participating teams. And this year it was not different - as many as 36 teams took part at IYPT Poland, two of them in the role of guest teams. This was, as in the previous few years, the highest number of participating teams ever. Some of the reasons may have been a relatively low travel expenses for countries that sometimes fall out due to financial difficulties, as well as Poland's long IYPT tradition promising a high quality of organisation. Finally, well running system of endorsers and guest teams and a lower participation fee made it easier for new countries to join IYPT for the first time. 

On the other hand, more teams and a smaller amount of resources made it harder for the organisers. The financial burden seamed to be really vexing and made the preparations rather hard. But in the end, all the core activities were secured smoothly and we were able to celebrate once again the same winner as the preceding years - the team from Singapore. 

The set of problems \cite{problems2019} was not less interesting than the previous year. Mechanics was represented by the \textit{Hurricane balls} task, about the well-known phenomenon of rotating pair of steel balls propelled by blowing through a tube. Although the effect was relatively well described by already published work \cite{hurricane}, teams still found a lot of topics that required further research. The basic principle of the raising pair of balls led to interesting discussions on forces, momenta and inertia, which were challenging for the students as well as for jurors. 

Another problem inspired by real world applications was \textit{Loud voices}, related to a passive megaphone. It turned out that a simple cone-shaped device is very effective in increasing the loudness of a narrow span of frequencies while a more sophisticated horn-shaped device seemed to be better for broad spectrum such as human voice. 
Interestingly, in both cases the effect of better transmittance was caused mostly by a better use of the source (loudspeaker or human) than by concentrating the energy into a specific direction. In other words, the overall power of the source was increased by using the device rather than the device would concentrate the power into a narrower angle. 

Fluid dynamics was at the core of the \textit{Funnel and Ball} problem, that asked the students to pick up a light ball by blowing air into a funnel. Here two basic hypotheses were presented - one relying on basic Bernoulli's principle of decreasing the pressure when increasing the velocity of the air by blowing. The other one was based on a more sophisticated (but much harder to quantify) Coanda effect causing the air to flow around the ball and concentrate under it leading to a local point of higher pressure. While the effect was experimentally observed, clear theoretical solution was not reached, making it a nice problem for future generation of students as well. 

Optics was represented by the \textit{Soy Sauce Optics} problem by introducing a thermal lens. Here it turned out that while a thin layer of soy sauce, if shined on with a strong laser, has very interesting optical properties, it can hardly by characterised as a simple lens. This lead again to heated discussions between reporting and opposing teams on the key aspects of the problem, as well as whether wave optics approach should have been used. 

\section{The Ring Oiler Problem}
\label{Oiler}

The problem statement reads as follows:

\textit{An oiled horizontal cylindrical shaft rotates around its axis at constant speed. Make a ring from a cardboard disc with the inner diameter roughly twice the diameter of the shaft and put the ring on the shaft. Depending on the tilt of the ring, it can travel along the shaft in either direction. Investigate the phenomenon.}

This problem is clearly of a pure mechanical nature - the only possible effects beyond might have been connected with the influence of oil and its surface tension, but it turned out experimentally that oil in fact is even not important for the experiment. The movement of the disk is very similar without the oil except the changed friction coefficient. In fact, it seems that the presence of the oil in the task was mostly motivated by the fact that such a device has been used practically to provide lubrication to moving shafts. This is also the reason why cardboard was mentioned in the task - it turned out that using a metal washer provided the same effect, while being much more symmetric and accurate in its dimensions.

\subsection{Basic movement and horizontal inclination of the ring}

\begin{figure}[t!]
    \centering
    \includegraphics[width = 0.8\textwidth]{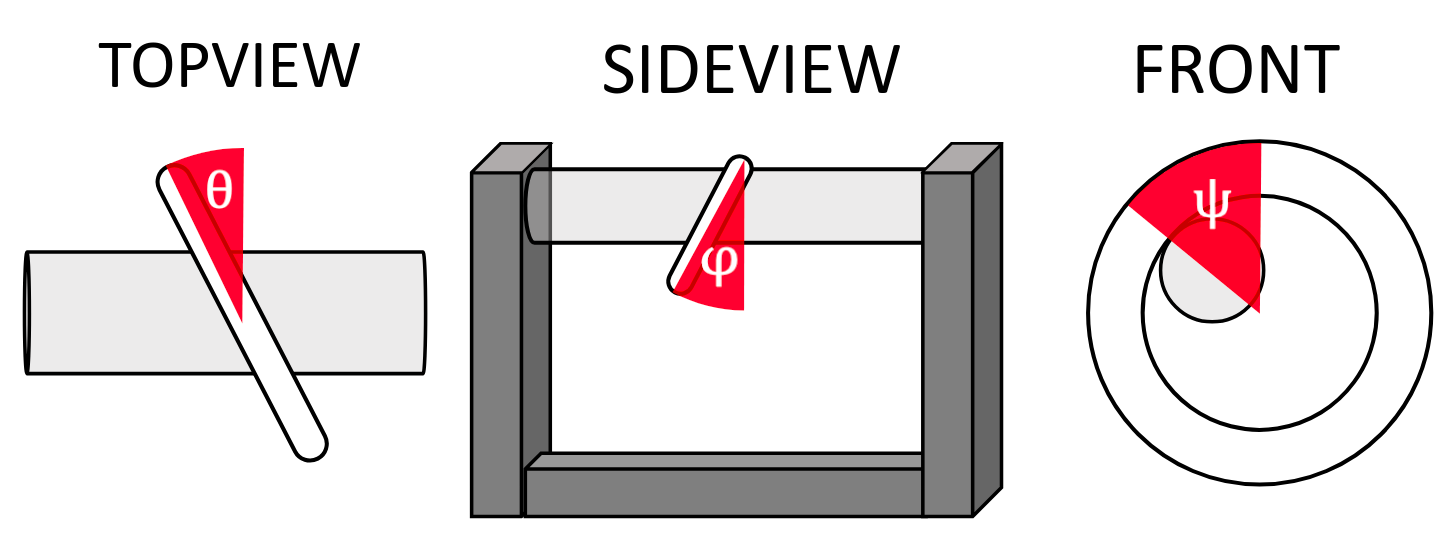}
    \caption{Definition of coordinates used in measurements. Angle $\theta$ is the tilt of the ring when viewed from the top, $\varphi$ describes the inclination from the vertical position and $\psi$ is the deflection of the center of mass to the side when viewed along the rod axis. At rest, all angles equal zero.}
    \label{fig:angles}
\end{figure}

One might very easily deduce that any possible steady state of the ring will
obey the non-slipping condition. Namely, the speed at the contact point will
be the same for the ring and the shaft. We can formalise this by stating
\[
\Omega\rho=\omega r,
\]
where $\rho$ is the inner radius of the ring, $r$ the radius of the shaft and
$\Omega$ and $\omega$ are the angular velocities of the shaft and the ring, respectively. One
steady state is clearly connected with the situation when the ring's axis stays in
line with the shaft and the ring rotates together with the shaft, without horizontal speed - however, this state is
unstable. After any fluctuation of the ring, characterised by an inclination
of its axis from the shaft's axis by and angle $\theta$ (as defined in Fig. \ref{fig:angles}) the ring starts to move
along the shaft. In such a case the non-slipping condition is more
complicated, as in fact the tangential velocity of the ring has to be equal to
the velocity of the shaft, namely
\begin{equation}
\Omega\rho=\omega r\cos(\theta),\label{theta}%
\end{equation}
making the ring accelerate when increasing its tilt. Equation (\ref{theta})
connects three experimentally accessible parameters, namely the frequency of
the shaft $\Omega$ and ring $\omega$, as well as the tilt of the ring $\theta$ (Fig.~\ref{fig:angles}).

\begin{figure}[h!]
    \begin{center}
    \includegraphics[width = 1.0\textwidth]{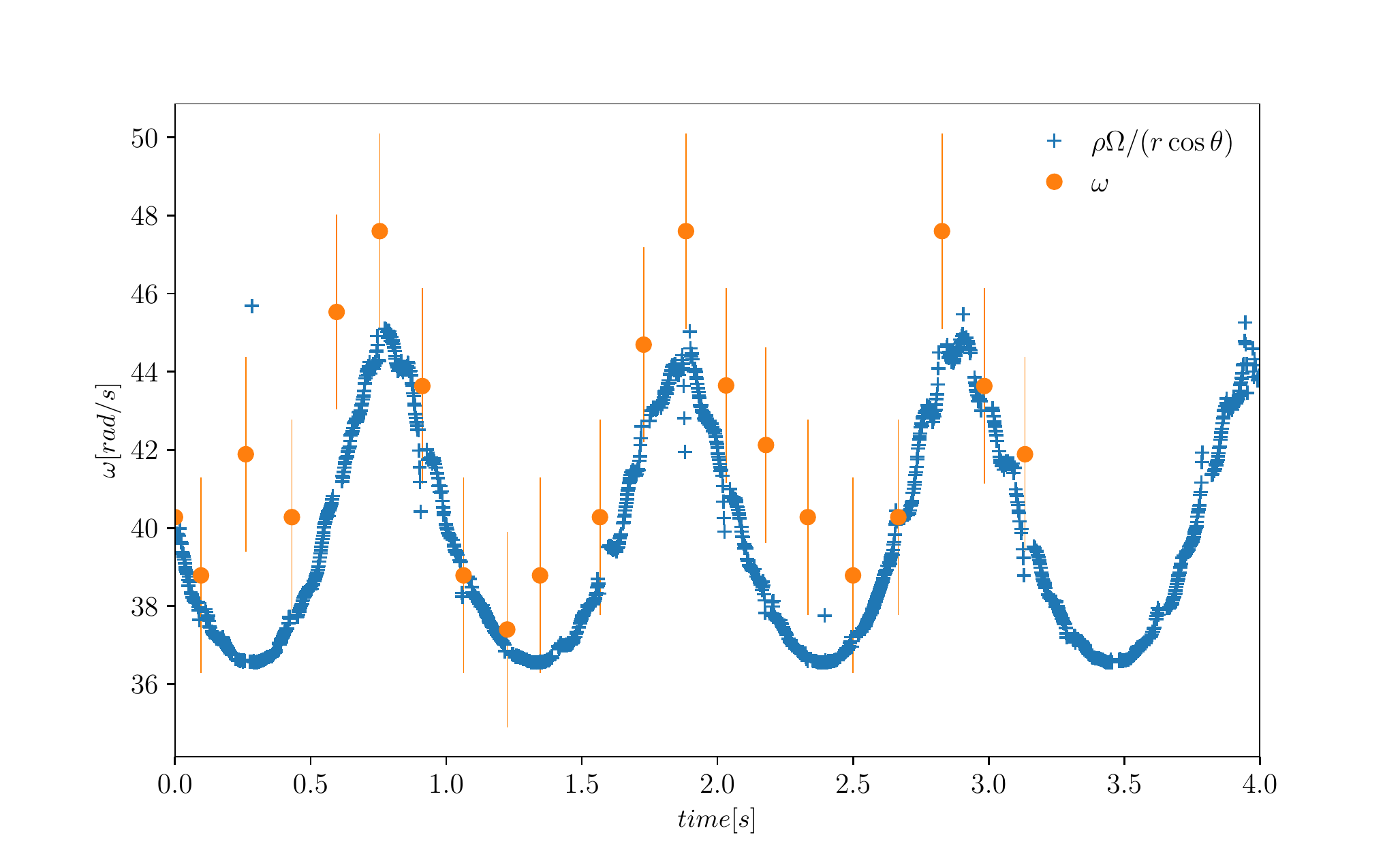}
    \end{center}
    \caption{Experimental dependence of the frequency of the ring $\omega$ and the
    tilt of the ring from the top view $\theta$. While there are differences of the order of the measurement precision, we see a clear correlation between the tilt and
    the speed of the ring.}%
    \label{fig_theta}%
\end{figure}

We performed an experiment with a metal shaft propelled by and AC 12/24V motor
controlled by a frequency controller. Its frequency was measured by a hall
sensor waked by a small magnet fixed on a ring at the end of the shaft. A
metal washer on the shaft was painted with a bright colour with a dot of a
contrasting colour on the side and recorded from three views - top, side and
along the shaft (front) by a high speed camera. A tracker software was used to determine
the position of the washer in each time as well as its rotational speed. In
Figure (\ref{fig_theta}) we see the experimental validation of Eq.~(\ref{theta}),
that is on the edge of confirming and disproving the theory.

\begin{figure}[ht]
    \begin{center}
    \includegraphics[width = .9\textwidth]{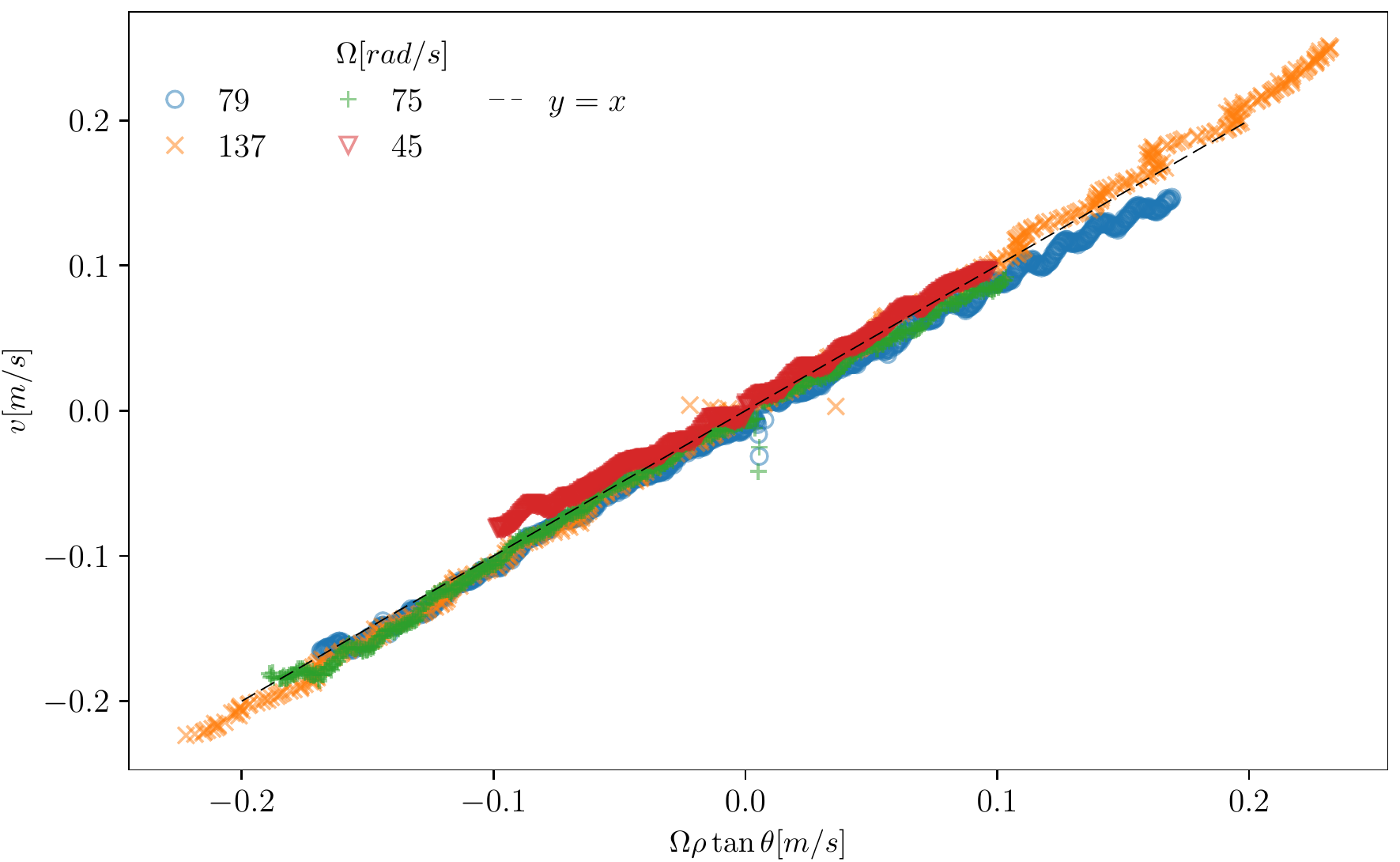}
    \end{center}
    \caption{Experimental test of validity of Eq.~(\ref{v}) for different values of
    $\Omega$. Here we see that the fit is almost perfect.}%
    \label{fig_v}%
\end{figure}

The washer, if tilted, moves along the shaft in the same way as a nut
moves along a screw. This allows us to derive an equation for the velocity of
the ring along the shaft
\begin{equation}
    v=\omega r\sin\theta=\Omega\rho\tan\theta.\label{v}%
\end{equation}
Eq.~(\ref{v}) can again be experimentally tested and the results are shown in
Fig. (\ref{fig_v}). Here we see that the fit is almost perfect for all tested
values of $\Omega$. 

\subsection{Vertical inclination}

As seen from the experiment, the ring will at least in the beginning
accelerate along the shaft - limits for this part of the motion will be
discussed in the next section. This acceleration will be induced by the
friction force on the contact point between the ring and the shaft. This
point, however, is distant from the center of mass of the shaft, so the
acceleration will cause a vertical inclination of the washer - the bottom part
will simply follow the upper part with some delay. In the limit of a slowly
rotating ring (where we just model it as a fixed object in an accelerated
frame) the vertical tilt angle of the ring (front view) is simply
\begin{equation}
    \tan\varphi=\frac{a}{g},\label{fi1}%
\end{equation}
where $a$\thinspace$=\overset{.}{v}$ is the acceleration of the ring along the
shaft and $g$ is the gravitational acceleration.

\begin{figure}[ht!]
    \begin{center}
        \includegraphics[width = .9\textwidth]{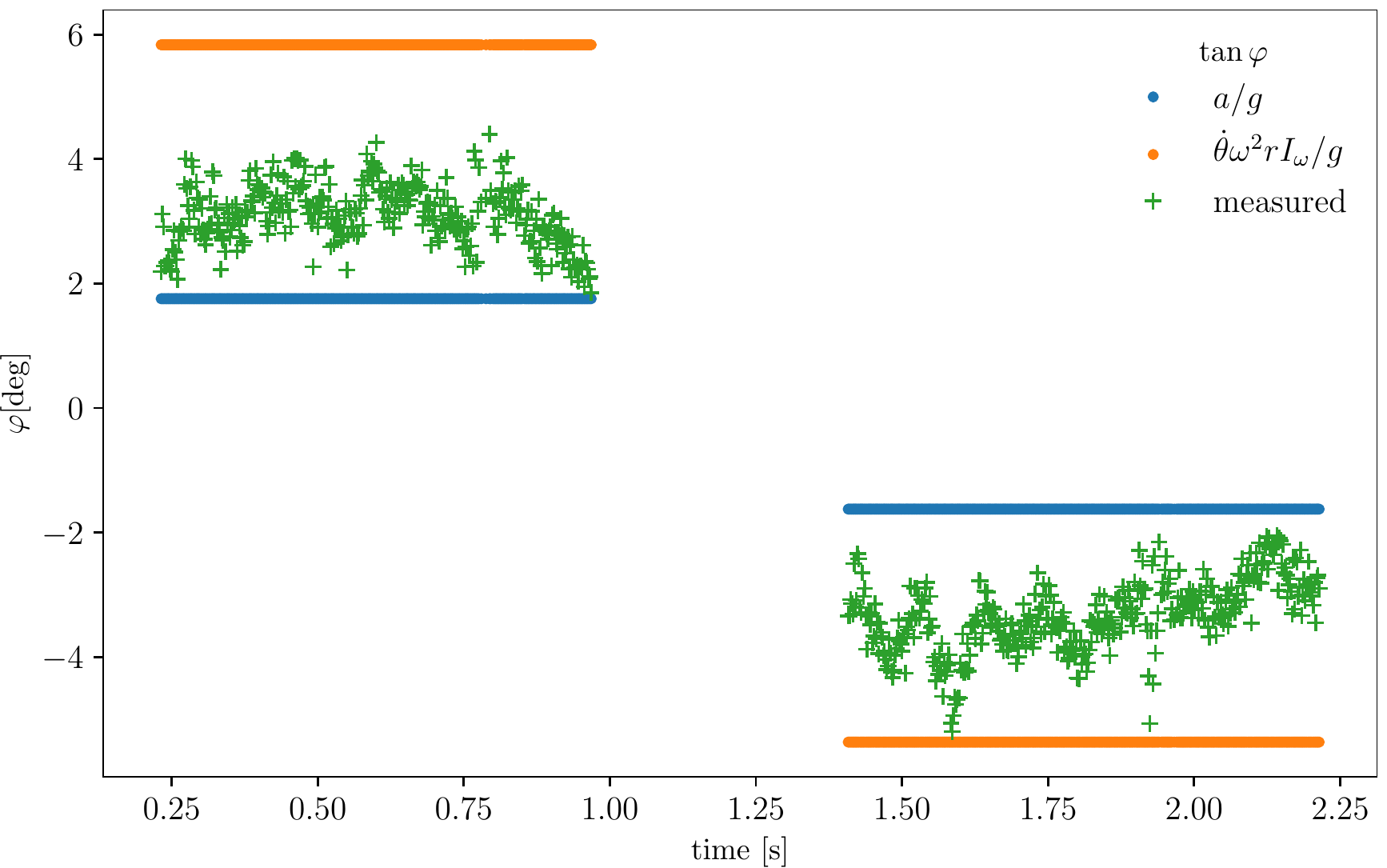}
         \includegraphics[width = .9\textwidth]{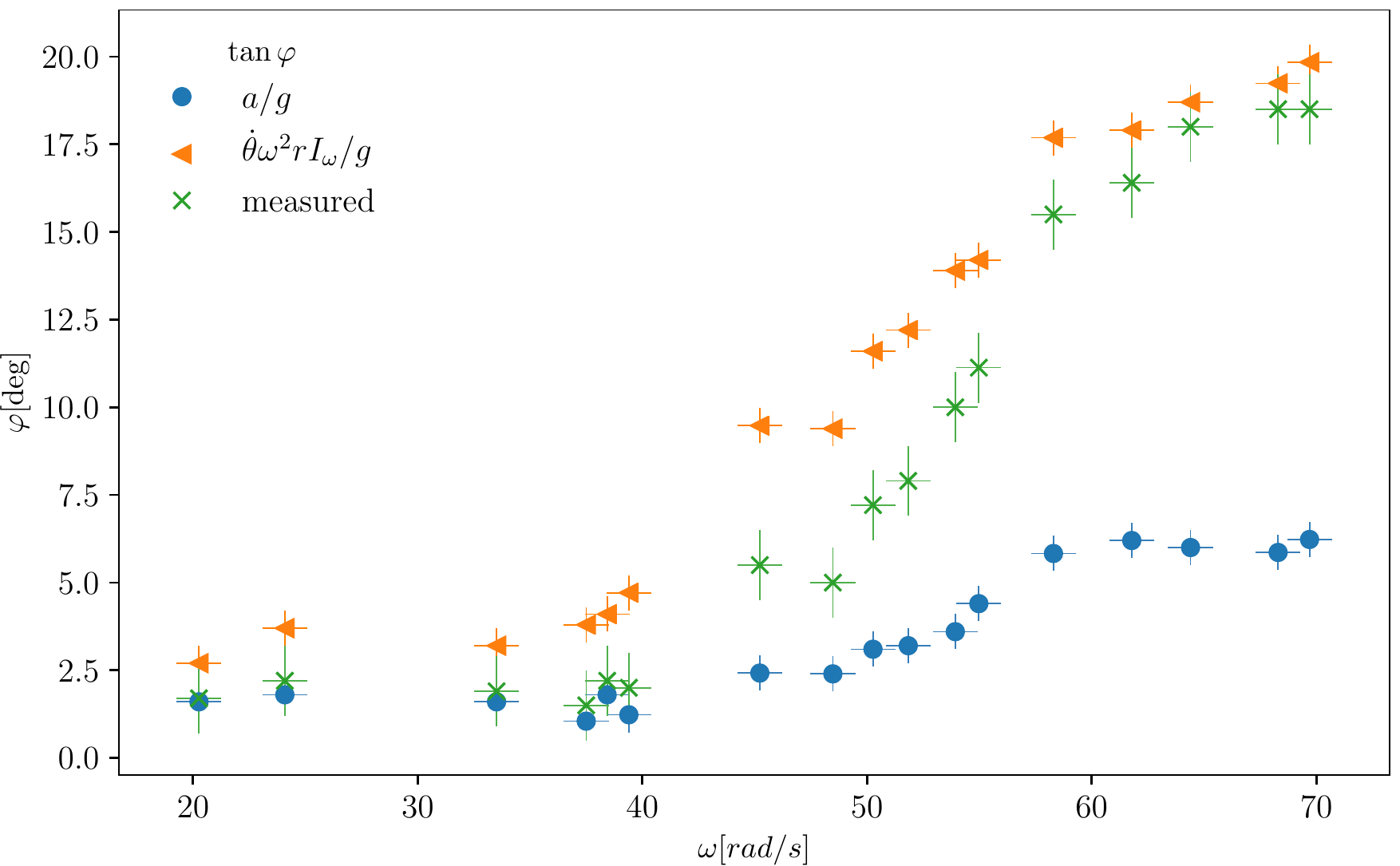}
    \end{center}
    \caption{We see that the experimental values are safely within the margins
given by the two limiting cases of non-rotating and fast-rotating ring. The upper graph represents one particular experiment, the bottom one aggregated value for different angular velocities of the shaft. We see that, expectantly, with increasing frequency the experimental value move from the stationary limit towards the high-speed limit.}%
    \label{fig_fi}%
\end{figure}

However, the ring rotates and thus has a nonzero angular momentum. In that case, the inertial force of 
the ring will cause a torque on the ring that will evolve the total angular momentum in a
non-trivial way. But we can still look at the limit of high angular momentum
(quickly rotating ring), where the torque will cause precession of the ring
while fixing the total angular momentum. This leads to a condition
\begin{equation}
    \tan\varphi=\frac{\overset{.}{\theta}\omega^{2}r I_{\omega}}{g},\label{fi2}%
\end{equation}
where $I_{\omega}$ is the angular momentum coefficient of the ring, depending on
the ratio of its inner and outer radius. One would expect that the
experimental values of the angle $\varphi$ will be somewhere in between the
limits given by Eqs.~(\ref{fi1}) and (\ref{fi2}). As we see in Fig.
(\ref{fig_fi}), this is indeed the case.

\subsection{Saturation}

\begin{figure}[ht!]
    \centering
    \includegraphics[width = 0.6\textwidth]{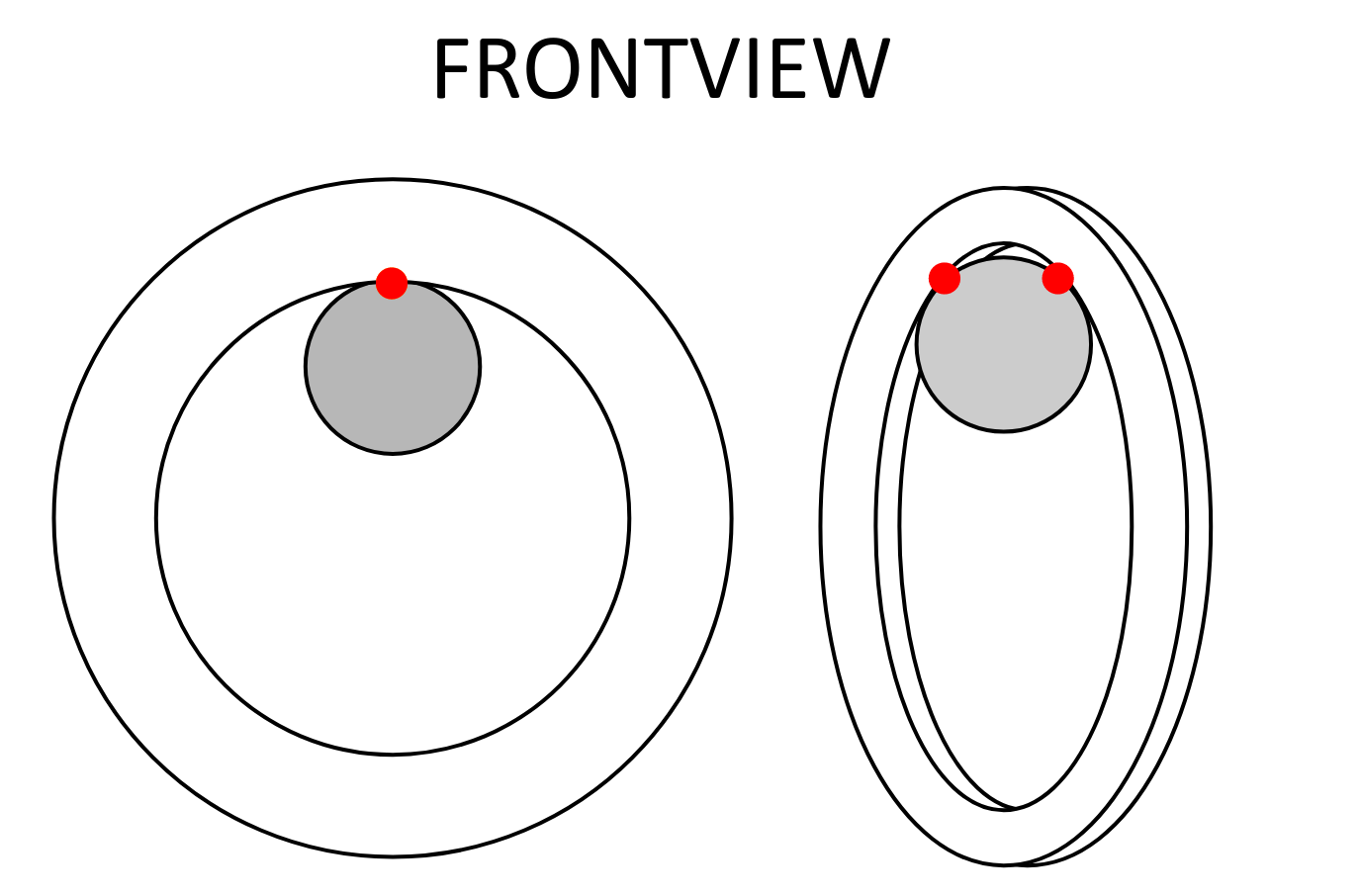}
    \caption{Geometrical limit for $\theta$ due to finite thickness of the shaft - saturation. When the angle $\theta$ reaches its maximum possible value $\theta = \theta_C$, it starts touching the rod in two contact points, what causes a torque acting against $\theta$. Depending on the elasticity of this "collision", the ring can either stay saturated for a perfectly inelastic collision with $\dot \theta = 0$ or, for a perfectly elastic collision, $\dot \theta$ changes its sign but not its absolute value. In our experiments, the situation close to the latter was observed.}
    \label{fig:saturation}
\end{figure}

\begin{figure}[ht!]
    \begin{center}
    \includegraphics[width = .9\textwidth]{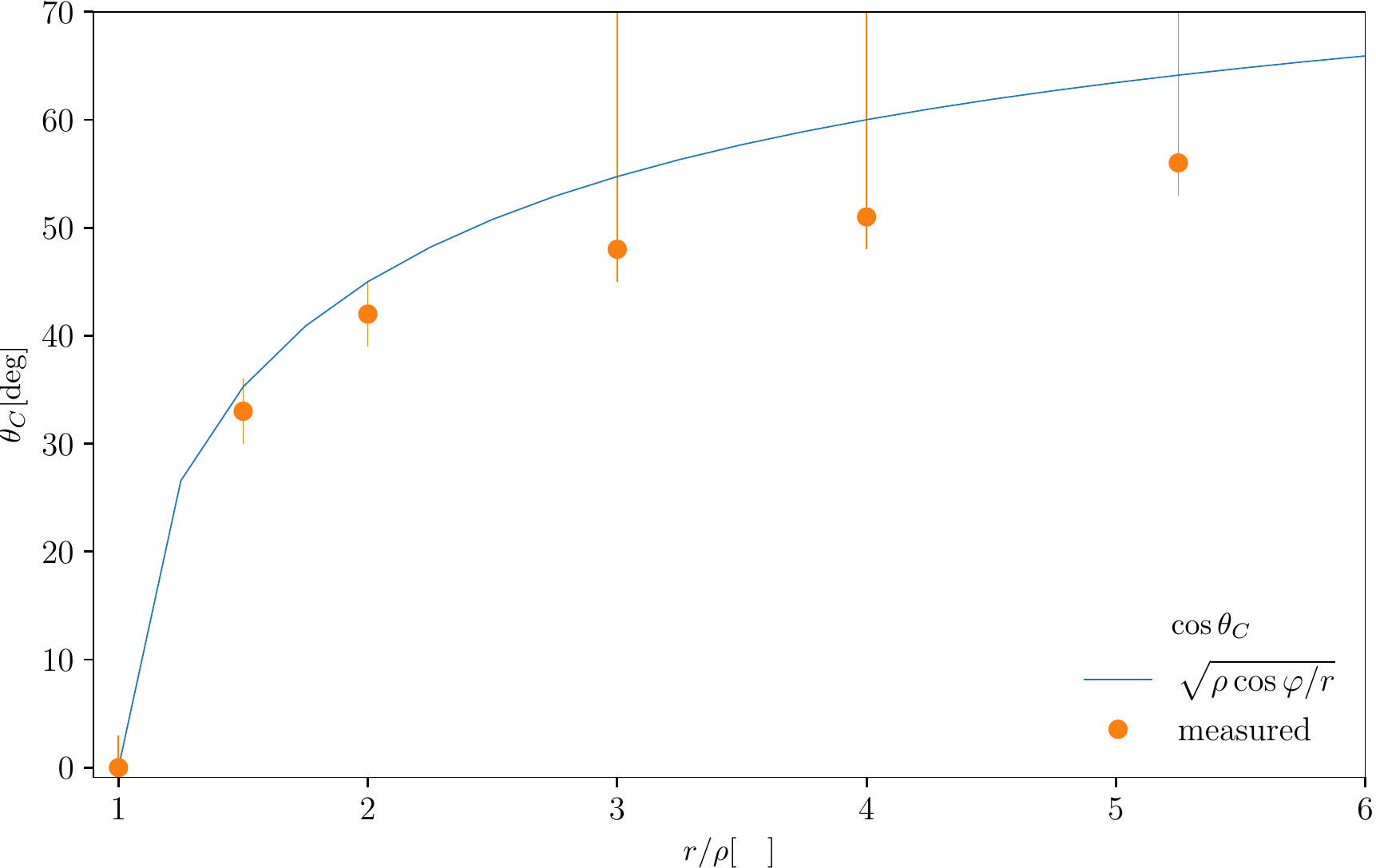}
    \end{center}
    \caption{Maximal angles $\theta$ reached for different ratios of the inner radius of the ring and radius of the rod compared with its theoretical expectations. For larger ratios (bigger rings) the saturation angles were not achieved before reaching the end of the rod, thus the experimental values form only a lower bound.}%
    \label{fig_saturation}%
\end{figure}

It is more than clear that the increase of the tilt angles has a limit - the
trivial one is $90$ $%
{{}^\circ}%
$, and a bit less trivial one is given by the fact that the shaft needs to pass
through the ring, limiting the maximal angle to
\begin{equation}
    r\cos\varphi=\rho.\label{tilt1}%
\end{equation}
This is, however, not the full story. We can look at the tilted ring from the
viewpoint along the shaft as an ellipse (see Fig.~\ref{fig:saturation}) with its axes given by the tilt
angles
\begin{align}
    a  & =r\cos\theta\\
    b  & =r\cos\varphi.
\end{align}
Condition (\ref{tilt1}) for both $\theta$ and $\varphi$ is limiting in terms of
fitting the shaft into the tilted ring. But before the angles reach this
threshold, another significant change happens on the contact point. Namely,
once the radius of the curvature at the contact point
\begin{equation}
    R=\frac{a^{2}}{b}=r\frac{\cos^{2}\theta}{\cos\varphi}\label{R}%
\end{equation}
gets smaller than the radius of the shaft $\rho$, the contact point
bifurcates into two points, as shown in Fig.~\ref{fig:saturation}. This will cause the ring to rise, but also the two
forces on the two contact points (summing up to the normal and friction force
as before) will have a non-zero torque acting against increasing the
horizontal tilt $\theta$. Thus we predict that maximal angle $\theta$ will be
limited by
\begin{equation}
    \cos\theta_{C1}=\sqrt{\frac{\rho}{r}\cos\varphi}.\label{theta_limit}%
\end{equation}
where $\theta_{C1}$ is the first geometrical limit.
And as we can see in the Figure (\ref{fig_saturation}), this is indeed the
case.

The answer to the question what happens after saturation is reached
depends on many aspects. Again, there exist two limiting cases. One is a
perfectly inelastic "collision" between the ring and the shaft. That is,
$\overset{.}{\theta}=0$ after reaching the saturation. This mostly happens in
case of using a soft (cardboard) disk and/or using thick oil on the shaft. In
that case, the ring stabilises in the motion along the shaft, $\varphi$
diminishes together with the acceleration and nothing interesting happens
till the ring collides with the end of the shaft.

Although above the limits set by the task itself, the more interesting case
is when the "collision" is close to elastic, what happens for the
experimentally examined case of hard shaft and washer ring with no oil used.
Here $\overset{.}{\theta}$ and $\varphi$ just basically change its sign with (almost)
keeping its absolute value (the washer flips). The resulting movement is very complex.

\subsection{Full movement}

So far we have mainly analysed the limits of the experimental data imposed by
some limiting theoretical predictions. While this is certainly more than
enough for an IYPT solution, especially in the case if a steady state is
reached in the presence of originally introduced cardboard disk with oil, it
is not a full explanation of the phenomenon. This is why we went further in
investigating the setup with steel washer without oil, which showed more
complexity by delivering almost perfectly elastic turnover of the disk when reaching the
saturation point. Here, the disk would oscillate from one side to the other
with an almost constant acceleration caused by the fact that both $\overset
{.}{\theta}$ and $\varphi$ are almost constant except during the turns (reflecting the fact that both
(\ref{fi1}) and (\ref{fi2}) lead to a constant $\varphi$ if $\overset{.}{\theta}$
and $a$ are constant).

In principle, as this is a purely mechanical system, it must have a full
solution hidden in the equations of motion. The complication is that even if
we neglect the air friction force and inaccuracies in the shape of the shaft
and the ring, the propelling force(s) (friction force(s) in the point(s)
between the ring and the shaft) are very complicated. Their direction and
magnitude depend both on the position and the speed of the disk relative to
the shaft (as friction force acts always against the relative velocity at the
contact point, unless the friction is static). This is why the motion can be
solved only numerically, not bringing any real insight into the reasons why
some of the parameters or their derivatives stay constant during most portion
of the movement. This is why we focused on some limiting cases again.

First we need to introduce a new parameter, namely the angle $\psi$. This
angle is defined as the deflection of the connection of the centre of mass of
the ring and the contact point (or center of the two contact points) from
vertical axis. From the physical point of view, this angle is connected with
the acceleration of the rotation speed of the ring.

\begin{figure}[ht!]
    \begin{center}
    \includegraphics[width = 1.0\textwidth]{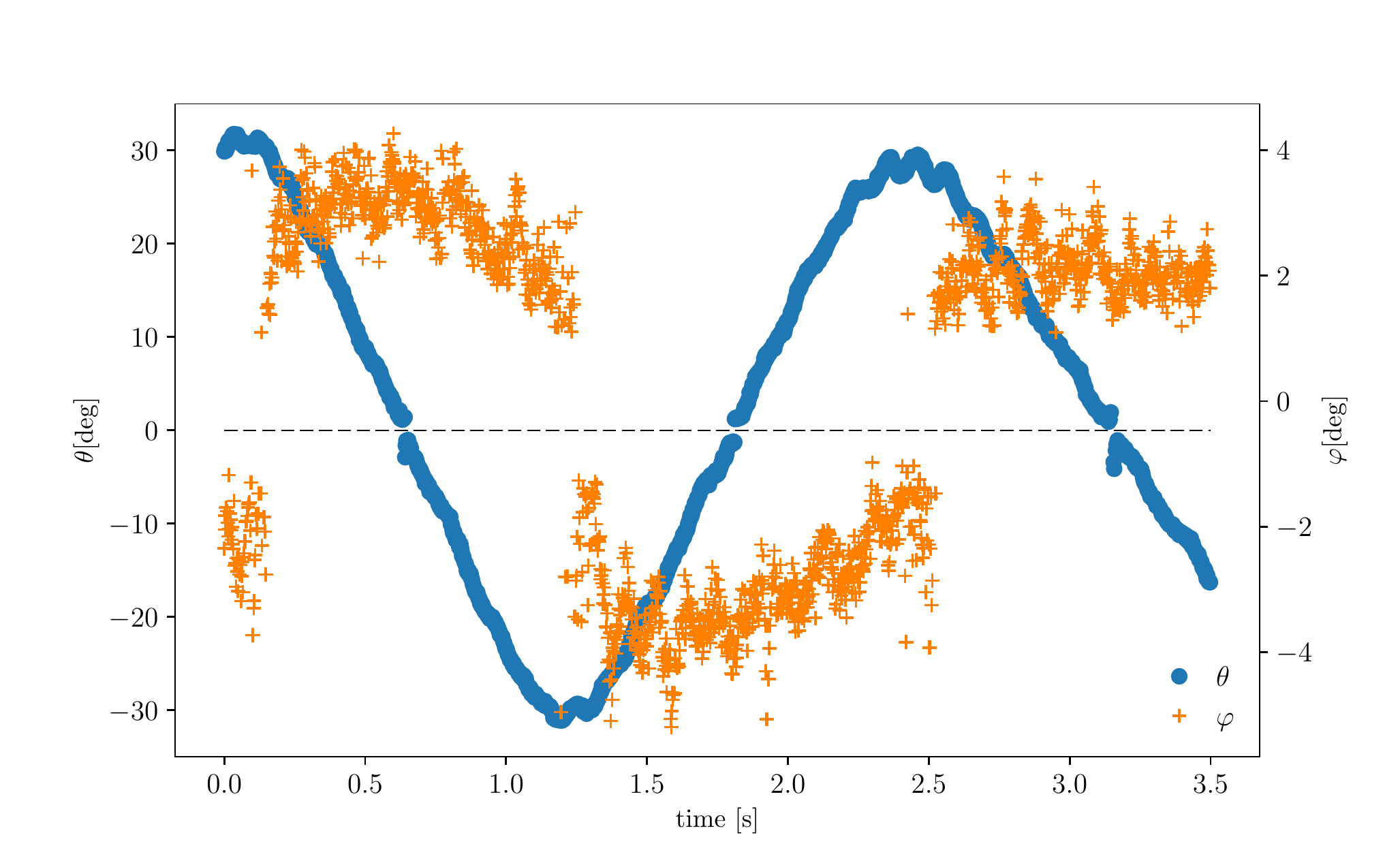}
    \end{center}
    \caption{Experimental result for slow rotations. We see that while $\varphi$ is
    more or less constant and quickly reverts on saturation points, $\theta$ has an
    almost constant derivative with an abrupt change of it at saturation connected with a
    few oscillations. Angle $\psi$ is not depicted, as it was negligible during
    the whole experiment. }%
    \label{fig_slow}%
\end{figure}

In the first limiting case $\psi=0$, expecting that the ring will
adapt its rotation speed almost immediately and its center of mass will be
exactly below the axis of the shaft. For small rotation frequencies the saturation turn
is also very short, so the reverse of the direction of the ring movement
happens almost instantaneously. Experimentally this was achieved for
frequencies of the shaft roughly below $10$ Hz. Even in this case the
equations of motion do not lead to a constant $\theta$, but rather to%
\begin{equation}
    \overset{.}{\theta}=-\frac{g}{\Omega\rho}\frac{\sin\varphi}{\left(1+\frac{R^2}{r^2}\right)-\frac{\cos\varphi
}{\cos^{2}\theta}}.\label{slow}%
\end{equation}
Eq.~(\ref{slow}) varies for constant $\varphi$ and the experimentally achieved
range of $\theta$ by about $30\%$, which is on the boundary of the
experimentally achieved precision, as shown in Fig. (\ref{fig_slow}).

\begin{figure}[ht!]
    \begin{center}
        \includegraphics[width = \textwidth]{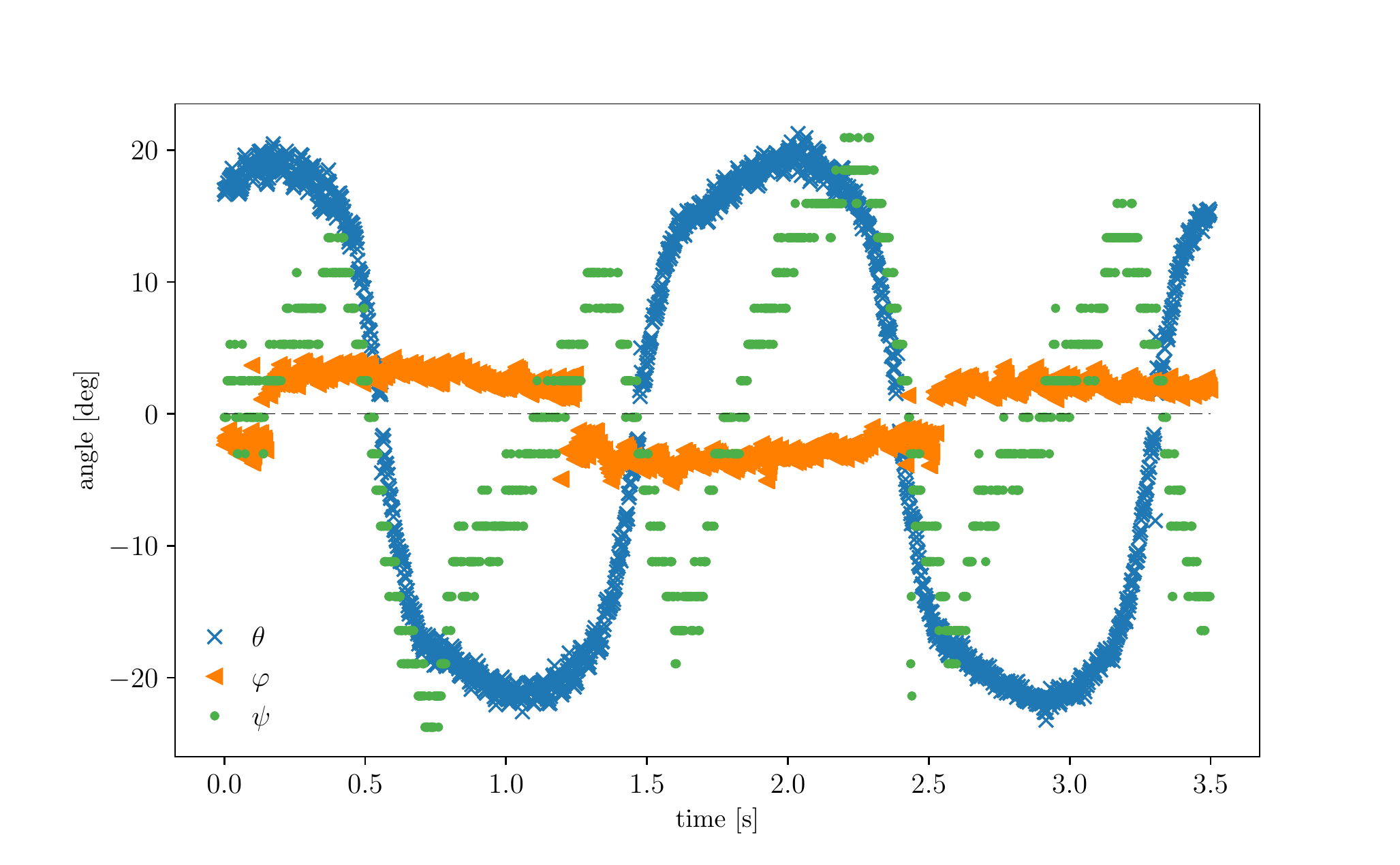}
    \end{center}
    \caption{Experimental result for fast rotations. Here two clear phases of the
    movement can be seen. One similar to the slow rotations, where $\varphi$ is
    almost constant while $\theta$ has an almost constant derivative. The other
    one is the change at saturation point, which is here smooth without any
    oscillations, but takes significant time. Most interestingly, angle $\psi$
    performs non-harmonic oscillations with double frequency comparing to $\theta$
    and $\psi$ - it has positive derivative while the disk is moving along the
    shaft and negative while at saturation. This is because the angular velocity of the disk is
    increasing while moving along the shaft and decelerating while reverting at saturation point.}%
    \label{fig_fast}%
\end{figure}

The other limiting case was when the rotation of the shaft was high,
experimentally achieved when $\Omega$ was much more than $10$ Hz. Here the
$\psi$ angle already achieved non-negligible values and the duration of the
"collision" at the saturation point was comparable to the time the disk was
moving at a constant speed. Unfortunately for this liming case the equations
of motion did not allow to make simplifications leading to an insight in the
phenomenon. This is somehow understandable when looking at the experimental
results depicted in Fig. (\ref{fig_fast}).   
Interestingly, the $\psi$ angle oscillates with a double frequency
compared to the other two angles. During the movement of the disk along the
shaft $\psi$ increases, while it decreases during the "collision" on the
saturation point. Interestingly, both of these changes seem to be close to
linear - the time derivative of $\psi$ is close to constant in each stage,
although it differs for the two stages both in sign and magnitude. There is no
more or less obvious reason for this deductible from the full set of equations
of motion.

To conclude, we have seen that even such a relatively simple mechanical
problem as Ring Oiler led to a very rich set of possibilities to point the
interest on. While the basic explanation of the movement of the ring along
the shaft is fairly simple, description of the whole phenomenon including
detailed understanding of the evolution of relevant parameters was not reached
even after a full scientific research conducted for more then a year.

\section{Conclusion}

The International Young Physicists' Tournament is much more than a competition for students. It is a unique event for a broad variety of target groups. High school students in the first place, of course, who will benefit from training both their hard skills (Physics, Maths and partly IT and Chemistry), but also in many soft skills like long term team work as well as presentation and discussion abilities. The number of participating countries keeps rising in the recent years in spite of the high entrance barrier caused by the fact that the level of preparation of the new teams must be comparable to the rest to allow a smooth discussion during the fights. This proves that many students (and also teachers, leaders and jurors) are still interested in complex activities rather than investing their time and effort into a series of short events with immediate gains. 

The set of problems for 2020 is ready \cite{problems2020}. Motivated students are asked to try to measure current using its heat effect, blow on a candle hidden behind a bottle or measure time using a Saxon Bowl. Levitation made it to the problem set again: this time, the levitating object will be a flea of a magnetic stirrer. Those using standard pencils while writing on paper can measure conductivity of the lines drawn and those extra creative can combine with the previous problem and do it via utilising the heat of the current produced. 
Others will enjoy trowing spinning playing cards to long distances and those who like physics of everyday life can come up with a way to pour salt or pepper most efficiently. 
Finally. the best out of the best will meet next year in July again at the 33rd International Young Physicists' Tournament 2020 in Timisoara.    

\section*{Acknowledgements}
This research was supported by the joint Czech-Austrian project MultiQUEST (I 3053-N27 and GF17-33780L), as well as project VEGA BeKvaK (2/0136/19).

\end{document}